\documentstyle[preprint,revtex]{aps}
\def\ltap{\raisebox{-.4ex}{\rlap{$\sim$}} \raisebox{.4ex}{$<$}}
\begin{document}
\thispagestyle{empty}
\font\fortssbx=cmssbx10 scaled \magstep2
\hbox{ 
\fortssbx University of Wisconsin Madison} 
\vspace{.3in}
\hfill\vbox{\hbox{\bf MAD/PH/730}
            \hbox{\bf RAL-92-072}
	    \hbox{November 1992}}\par
\vspace{.2in}
\begin{title}
Implications of $b\rightarrow s\gamma $ decay measurements in testing \\
the MSSM Higgs sector
\end{title}
\author{V.~Barger,$^*$  M.~S.~Berger,$^*$ and
R.~J.~N.~Phillips$^\dagger$}
\begin{instit}
$^*$Physics Department, University of Wisconsin, Madison, WI 53706, USA\\
$^{\dagger}$Rutherford Appleton Laboratory, Chilton, Didcot, Oxon OX11  0QX,
England
\end{instit}
\begin{abstract}
\baselineskip=20pt 
\nonum\section{abstract}
The observation that the $b \rightarrow s\gamma $ decay rate is close to the
Standard Model value implies a large mass for the charged-Higgs boson
in the Minimum Supersymmetric Standard Model, that nearly closes the
$t \rightarrow b H^+$  decay channel.  For $m_t = 150$ GeV, the parameter
region $m_A \ltap 130$ GeV is excluded; this largely pre-empts
LEP II searches and also partially excludes a region that would be
inaccessible to MSSM Higgs boson searches both at LEP II and at SSC/LHC.
\end{abstract}

\newpage
   In the Minimum Supersymmetric Standard Model (MSSM), the inclusive
$b \rightarrow s \gamma$  branching fraction is very sensitive to charged Higgs
boson loop contributions,  since they interfere constructively with
the Standard Model (SM) $W$-loop amplitude and both receive a strong
QCD enhancement\cite{bsg,bhp}.  The present experimental upper
limit  $B(b \rightarrow s \gamma) < 8.4\times 10^{-4}$
from the
CLEO collaboration\cite{cleo} already implies
severe constraints\cite{jh} on the charged Higgs boson mass
$m_{H^{\pm }}$\cite{bhp}
for given top quark mass $m_t$ and
model parameter $\tan \beta = v_2/v_1$, the ratio of the two vacuum
expectation values appearing in the MSSM. In the present Letter we point out
that these constraints almost
close the interesting decay channel  $t \rightarrow b H^+$ (the basis of all
viable  $H^{\pm }$  studies at hadron
colliders\cite{exp,kz,baer,gunion,bcps})
and exclude most of the  ($m_A, \tan \beta$) parameter region
accessible to LEP II Higgs boson searches.  They also partially
exclude a parameter
region believed to be inaccessible to combined LEPI, LEP II and
SSC/LHC searches\cite{kz,baer,gunion,bcps}.  These important constraints will
become even more far-reaching when more precise theoretical calculations are
made and a more accurate determination of
$B(b \rightarrow s \gamma)$  becomes possible,
for example at future $B$-factories.

   For calculating QCD enhancements of the $b\rightarrow s\gamma $ decay
amplitudes, we use the prescription of
Grinstein, Springer and Wise\cite{gsw}. The relevant operator
arising from the dominant  $tW^+$  and $tH^+$  loop contributions at scale
$m_b$ has the form
\begin{eqnarray}
c_7(m_b)=\left [{{\alpha _s(M_W)}\over {\alpha _s(m_b)}}\right ]^{16/23}
\Bigg \{c_7(M_W) &-& {8\over 3} c_8(M_W)\left [1-\left (
{{\alpha _s(m_b)}\over {\alpha _s(M_W)}}\right )^{2/23}\right ]
\nonumber \\
&+&\left . {232\over 513}
\left [1-\left (
{{\alpha _s(m_b)}\over {\alpha _s(M_W)}}\right )^{19/23}\right ]
\right \} \;, \label{c7}
\end{eqnarray}
where for the MSSM
\begin{eqnarray}
   c_7(M_W) &=& -{1\over 2} A(x) - B(y) - {1\over {6\tan^2 \beta}} A(y)
\;, \label{c7II} \\
   c_8(M_W) &=& -{1\over 2} D(x) - E(y) - {1\over {6\tan^2 \beta}} D(y)
\;, \label{c8II}
\end{eqnarray}
with  $x = (m_t/M_W)^2$, $y = (m_t/m_{H^{\pm}})^2$. The functions $A$, $B$,
$D$ and $E$
are defined in Ref.~\cite{gsw}.  The ratio of
$\Gamma (b \rightarrow s \gamma)$ to
the inclusive semileptonic decay width is then given by
\begin{equation}
{{\Gamma(b\rightarrow s \gamma)}\over {\Gamma (b\rightarrow ce\nu )}}
={{6\alpha }\over {\pi \rho \lambda }}|c_7(m_b)|^2 \;, \label{ratio}
\end{equation}
where $\alpha $ is the electromagnetic coupling.   The phase-space factor
$\rho $ and the QCD correction factor $\lambda $ for the semileptonic process
are given by $\rho = 1 - 8r^2 + 8r^6 - r^8 - 24r^4\ln(r)$ with $r= m_c/
m_b$  and  $\lambda = 1 - {2\over 3} f(r,0,0)\alpha_s^{}(m_b)/\pi $
with $f(r,0,0)=2.41$\cite{QCDcorr}.
Note that the $m_b^5$  dependence of the partial widths
cancels out in Eq.~(\ref{ratio}),
and also the
CKM matrix elements cancel to a good approximation. We ignore charm
quark contributions ($\sim 0.1$\% in the amplitude). We evaluate the
$b \rightarrow s \gamma $ branching fraction from Eq.~(\ref{ratio})
using the accurately
determined semileptonic branching fraction $B(b \rightarrow c e \nu) = 0.107$
and the estimate $\alpha _s(M_W)/\alpha _s(m_b) = 0.548$ based\cite{bbo}
on a three-loop formula with $m_b=4.25$ GeV.

The $B(b\rightarrow s\gamma )$
results depend sensitively on both  $m_{H^{\pm }}$ and $\tan \beta$. The
MSSM sets a lower bound  $m_{H^{\pm }}^2 = M_W^2 + m_A^2 > M_W^2$
at tree level; with
one-loop radiative corrections for $M_{SUSY}^{}=1$ TeV
and experimental limits on $m_A>40$ GeV this
bound becomes approximately $m_{H^{\pm }} > 90$ GeV, well above the LEP
detection limits for $H^{\pm }$.  There are bounds  $m_t/600 < \tan \beta <
600/m_b$  from requiring Yukawa couplings to remain perturbative\cite{bhp}
and  $\tan \beta < 85$  from the proton lifetime\cite{hmy}.
There are also constraints from low-energy data (principally
$B-\overline{B}$, $D-\overline{D}$, $K-\overline{K}$
mixing) that exclude low values of
$\tan \beta $\cite{bhp,buras} but these are less stringent than the
$b \rightarrow s \gamma $  constraint of present concern.

   Figure 1 shows the dependence of $B(b \rightarrow s \gamma)$ on
$\tan \beta $
for $m_t = 150$ GeV, with various choices of $m_{H^{\pm }}$.  We
see that the CLEO bound  $B < 8.4 \times 10^{-4}$  not only excludes small
values
of $\tan \beta$ for any $m_{H^{\pm }}$, but also excludes a range of lower
$m_{H^{\pm }}$ values ($m_{H^{\pm }}\ltap 155$ GeV in the
$m_t=150$ GeV case shown) for any $\tan \beta$.  However this lower
limit on $m_{H^{\pm }}$ depends quite sensitively on the theoretical
calculation as discussed below.

   Figure 2 translates the  $b \rightarrow s \gamma$  bound  into the
$(m_{H^{\pm }},
\tan \beta)$ plane for $m_t=150$ GeV
and compares it with the bounds from perturbativity,
proton decay and the MSSM $m_{H^{\pm }}$ formula.  The bound is beyond the
threshold $m_{H^{\pm }} = m_t - m_b$, so that the
decay mode $t \rightarrow b H^+$ is
closed.
As $m_t$ is reduced, the $b\rightarrow s \gamma $ bound and the
$t\rightarrow bH^+$ threshold move left toward
the MSSM
constraint on $m_{H^{\pm }}$ (that we have calculated from the one-loop mass
formula\cite{brig} plus LEP limits on $m_A$\cite{LEP}).
For $m_t\ltap 130$ GeV the $b\rightarrow s \gamma $ bound overtakes the
$t\rightarrow b H^+$ threshold and this decay mode becomes marginally open,
in our present calculations. At $m_t\simeq 95$ GeV, the threshold crosses the
MSSM bound and the $t\rightarrow bH^+$ decay becomes closed once more.

   Figure 3 translates the  $b \rightarrow s \gamma $ bound into the
$(m_A,tan\beta)$ plane, where coverage of the MSSM is usually
discussed\cite{kz,baer,gunion,bcps}.  The area below and to the left of
the $b\rightarrow s \gamma $ curve
is excluded.  The boundary of the region accessible to  $e^+e^- \rightarrow
Zh,Ah$  searches at LEP II is below and to the left of the
dashed curve;  we see that a
large part of this LEP II range is pre-empted.   Heavy shading shows the
area of parameter space that appears to be inaccessible to MSSM
Higgs boson searches at LEP I, LEP II and SSC/LHC (reproduced here from
Ref.~\cite{bcps}); we see that this inaccessible region is already partially
covered by the $b \rightarrow s \gamma $ bound.
With further improvements in measurements of $B(b \rightarrow s \gamma )$,
the coverage of the MSSM may in fact be complete after all.

The $b\rightarrow s \gamma $ bound of Fig. 3 also leads to an interesting
possible correlation between $m_t$ and the lighter CP-even Higgs boson mass
$m_h$, if  we inject an additional theoretical requirement on Yukawa couplings
$\lambda _b(M_G^{})=\lambda _{\tau}(M_G^{})$ from SUSY-GUT
unification following Refs.~\cite{bbo,yukawa}.
Given $m_t < 175$ GeV, the authors of Ref.~\cite{bbo} find just
two solutions for $\beta $, namely $\sin \beta=0.78(m_t/150 {\rm GeV})$ and
$\tan \beta > m_t/m_b$. For $m_t=150$ GeV, the first solution gives
$\tan \beta=1.25$ for which the $b\rightarrow s\gamma$ bound allows only the
limited range $68<m_h<76$ GeV (assuming a squark mass scale of order 1 TeV as
in Ref.~\cite{bcps}; this range shifts up (down) by approximately 10 GeV when
the SUSY scale is increased (decreased) by a factor two);
such an observed correlation of $m_h$ with $m_t$ would support this solution.
Note that
threshold corrections to the Yukawa coupling unification constraint could
shift this predicted Higgs mass range somewhat. For
the large-$\tan \beta$ solution, the $b\rightarrow s\gamma$ bound
does not effectively constrain $m_h$.

The above analysis has neglected various theoretical uncertainties. The
leading-log form in Eq.~(\ref{c7}) is obtained by truncating the
anomalous dimension matrix to three operators. This leads to an uncertainty of
at most 15\%\cite{gsw} to $c_7^{}(m_b)$ in the Standard Model.
This uncertainty should be
reduced somewhat in the MSSM where the charged Higgs contribution increases
the leading order result as in Eqs.~(\ref{c7II})-(\ref{c8II}).
The next-to-leading-log corrections are expected to be
about 20\% of the leading-log corrections in the decay rate. Therefore
these two effects together, if additive,
could lead to an overall uncertainty in the decay
rate of up to 50\% which we have not included in our analysis.
The lower limit on the charged Higgs mass is sensitive to
these uncertainties as can be seen in Figure 1. Thus if the theoretical
uncertainties mentioned above could be reduced, the limits on the charged Higgs
boson mass would be more reliable.
The leading-log result in Eq.~(\ref{c7}) is
obtained by integrating out the top quark (and the
charged Higgs boson) at the mass
of the $W$. Recently the leading corrections have been obtained when the
top quark is integrated out at a scale larger than $M_W^{}$\cite{cg}. These
corrections enhance the decay rate by as much as 14\%.
This enhancement may partially compensate for the
theoretical uncertainties enumerated above.

We have neglected any other sources of FCNC in the
supersymmetric model to the
$b\rightarrow s \gamma $ decay rate\cite{bbmr}.
Only if extra contributions conspire to reduce
the decay rate can the bounds in this
paper be evaded.

\acknowledgements
We thank J.~L.~Hewett and T.~Rizzo for discussions.
This research was supported
in part by the University of Wisconsin Research Committee with funds granted by
the Wisconsin Alumni Research Foundation, in part by the U.S.~Department of
Energy under contract no.~DE-AC02-76ER00881, and in part by the Texas National
Laboratory Research Commission under grant no.~RGFY9173.

\newpage
\nonum\section{figures}
\vskip 0.5in

\noindent Fig. 1. Calculated dependence of the inclusive branching fraction
   $B(b \rightarrow s\gamma )$ on $\tan \beta$ in the MSSM,
 with various values of
   $m_{H^{\pm }}$ , for $m_t = 150$ GeV.
   The shaded area is excluded by the CLEO bound
   $B(b\rightarrow s\gamma ) < 8.4\times 10^{-4}$.
\vskip 0.5in

\noindent Fig. 2. Comparison of the $b\rightarrow s\gamma$ bound with
other bounds from
   perturbativity, proton lifetime and the MSSM mass formulas,
   in the ($m_{H^{\pm }}, tan\beta$) plane, for
   $m_t=150$ GeV.  The threshold for $t\rightarrow bH^+$ decay at
   $m_{H^{\pm }} = m_t - m_b$ is also shown. Shaded areas are excluded by
   one or more bounds.
\vskip 0.5in

\noindent Fig. 3. Comparison of the $b \rightarrow s \gamma $ bound in the
   ($m_A, \tan \beta$)
   plane, with the region accessible to LEP I and LEP II searches
   (lightly shaded) and the
   region apparently inaccessible to LEP I, LEP II and SSC/LHC
   MSSM Higgs boson searches (heavily shaded vertical region)
   for $m_t=150$ GeV.


\begin{references}

\bibitem{bsg} T.~G.~Rizzo, Phys.\ Rev.\ {\bf D38}, 820 (1988); B.~Grinstein and
M.~B.~Wise, Phys.\ Lett.\ {\bf B201}, 274 (1988); R.~Grigjanis,
P.~J.~O'Donnell, M.~Sutherland and H.~Navelet, Phys.\ Lett.\ {\bf B213}, 355
(1988), {\bf B286} 413 (1992); W.~-S.~Hou and R.~S.~Willey,
ibid {\bf B202}, 591 (1988); T.~D.~Nguyen et al., Phys.\ Rev.\ {\bf D37},
3220 (1988); D.~Ciuchini,
Mod.\ Phys.\ Lett.\  {\bf A4}, 1945 (1989); J.~L.~Hewett et al., Phys.\ Rev.\
{\bf D39}, 250 (1989).

\bibitem{bhp} V.~Barger, J.~L.~Hewett and R.~J.~N.~Phillips, Phys.\ Rev.\
   {\bf D41}, 3421 (1990).

\bibitem{cleo} CLEO collaboration: report by D.~Kreinick at the Carleton
   {\it Beyond the Standard Model Conference}, Ottawa, Canada, June 1992.

\bibitem{jh} J.~L.~Hewett and R.~J.~N.~Phillips, Talks given at the
   {\it Top Quark Workshop on High-Energy
   Physics with Colliding Beams}, Madison, Wisconsin, November 1992.

\bibitem{exp} See Proceedings of 1990 Snowmass Summer Study on High
   Energy Physics; Proceedings of Large Hadron Collider Workshop
   at Aachen 1990, CERN 90-10; GEM Letter of Intent, B.~Barish,
   W.~Willis et al., GEM TN-92-49; SDC Technical Design Report,
   SDC-92-201.

\bibitem{kz} Z.~Kunszt and F.~Zwirner, CERN-TH.6150 (1991).

\bibitem{baer} H.~Baer, M.~Bisset, C.~Kao, and X.~Tata,
   Phys.\ Rev.\ {\bf D46}, 1067 (1992);
   H.~Baer, M.~Bisset, D.~Dicus, C.~Kao, and X.~Tata, Florida State report
   FSU-HEP-920724 (1992).

\bibitem{gunion} J.~F.~Gunion et al., Phys.\ Rev.\ {\bf D46}, 2040 (1992);
   J.~F.~Gunion and L.~H.~Orr, ibid {\bf D46}, 2052 (1992),
   J.~F.~Gunion, H.~E.~Haber and
   C.~Kao, ibid {\bf D46}, 2907 (1992); R.~M.~Barnett et al., UC-Davis report
   UCD-92-14 (1992).

\bibitem{bcps} V.~Barger, K.~Cheung, R.~J.~N.~Phillips and A.~L.~Stange,
   UW Madison report MAD/PH/696 (1992); V.~Barger, M.~S.~Berger,
   R.~J.~N.~Phillips and A.~L.~Stange, Phys.\ Rev.\ {\bf D45}, 4128 (1992).

\bibitem{gsw} B.~Grinstein, R.~Springer and M.~B.~Wise, Nucl.\ Phys.\
   {\bf B339}, 269 (1989).

\bibitem{QCDcorr} N.~Cabibbo and L.~Maiani, Phys.\ Lett.\ {\bf B79},
109 (1978);
M.~Suzuki, Nucl.\ Phys.\ {\bf B145}, 420 (1978);
N.~Cabibbo, G.~Corb\`{o}, and  L.~Maiani, ibid {\bf B155}, 93 (1979).

\bibitem{bbo} V.~Barger, M.~S.~Berger and P.~Ohmann, UW-Madison report
   MAD/PH/711 (1992).

\bibitem{hmy} J.~Hisano, H.~Murayama and T.~Yanagida, Tohoku University
   report TU-400 (1992)

\bibitem{buras} A.~J.~Buras et al., Nucl.\ Phys.\ {\bf B337}, 284 (1990).

\bibitem{brig} A.~Brignole et al., Phys.\ Lett.\ {\bf B271}, 123 (1991).

\bibitem{LEP} ALEPH collaboration, Phys.\ Lett.\ {\bf B265}, 475 (1991);
   DELPHI collaboration, ibid {\bf B245}, 276 (1990);
L3 collaboration, ibid {\bf B251}, 311
   (1990); OPAL collaboration, Z.\ Phys.\ {\bf C49}, 1 (1991).

\bibitem{yukawa} S.~Kelley, J.~L.~Lopez and D.~V.~Nanopoulos, Phys.\ Lett.\
{\bf B274}, 387 (1992); J.~Ellis, S.~Kelley, D.~V.~Nanopoulos, Nucl.\ Phys.\
{\bf B373}, 55 (1992); H.~Arason, D.~J.~Casta\~{n}o, B.~Keszthelyi,
S.~Mikaelian,
E.~J.~Piard, P.~Ramond and B.~D.~Wright, Phys.\ Rev.\ Lett.\
{\bf67}, 2933 (1991), P.~Ramond, University of Florida preprint UFIFT-92-4;
H.~Arason, D.~J.~Casta\~{n}o, E.~J.~Piard and P.~Ramond,
University of Florida preprint UFIFT-92-8;
G.~Anderson, S.~Dimopoulos, L.~J.~Hall and S.~Raby, Ohio State
preprint OHSTPY-HEP-92-018 (1992).

\bibitem{cg} P.~Cho and B.~Grinstein, Nucl.\ Phys.\
   {\bf B365}, 279 (1991).

\bibitem{bbmr} S.~Bertolini, F.~Borzumati, and A.~Masiero,
   Nucl.\ Phys.\ {\bf B294}, 321 (1987);
   S.~Bertolini, F.~Borzumati, A.~Masiero and G.~Ridolfi,
   Nucl.\ Phys.\ {\bf B353}, 591 (1991).

\end{references}
\end{document}